\documentclass[aps,prl,reprint,amssymb,showpacs,twocolumns,superscriptaddress,notitlepage,longbibliography]{revtex4-2} 
\usepackage{bm}
\usepackage{graphicx}
\usepackage{dcolumn}
\usepackage{braket}
\usepackage{bbold}
\usepackage{natbib}
\usepackage{amsmath}
\usepackage{color}
\usepackage{dsfont}
\usepackage{comment}
\usepackage{ragged2e}
\usepackage{tikz}
\usepackage{adjustbox}

\usepackage[normalem]{ulem}

\definecolor{mayablue}{rgb}{0.45, 0.76, 0.98}
\definecolor{deepskyblue}{rgb}{0.0, 0.75, 1.0}
\definecolor{dodgerblue}{rgb}{0.12, 0.56, 1.0}
\definecolor{ultramarineblue}{rgb}{0.25, 0.4, 0.96}
\definecolor{portlandorange}{rgb}{1.0, 0.35, 0.21}
\definecolor{purple(x11)}{rgb}{0.63, 0.36, 0.94}
\definecolor{gold}{rgb}{0.99, 0.76, 0.0}

\definecolor{forestgreen}{RGB}{34,139,34}

\usepackage[colorlinks]{hyperref}
\hypersetup{   
	linkcolor=black, 
	urlcolor=blue,
    citecolor=blue,
	pdfstartview=
}
\usepackage{orcidlink}

\usepackage[mathscr]{euscript}

\usepackage{lipsum}

\begin{document}

\title{
{Reliable quantum advantage in quantum battery charging}
}

\author{Davide Rinaldi\,\orcidlink{0009-0002-2562-0807}} 
\affiliation{Dipartimento di Fisica ``A. Volta,'' Universit\`a di Pavia, via Bassi 6, 27100 Pavia (Italy)}

\author{Radim Filip\,\orcidlink{0000-0003-4114-6068}}
\affiliation{Department of Optics and Quantum Optics Laboratory, Palacký University, 17. listopadu 12, 77146 Olomouc (Czech Republic)}

\author{Dario Gerace\,\orcidlink{0000-0002-7442-125X}}
\affiliation{Dipartimento di Fisica ``A. Volta,'' Universit\`a di Pavia, via Bassi 6, 27100 Pavia (Italy)}

\author{Giacomo Guarnieri\,\orcidlink{0000-0002-4270-3738}}
\affiliation{Dipartimento di Fisica ``A. Volta,'' Universit\`a di Pavia, via Bassi 6, 27100 Pavia (Italy)}

\begin{abstract}
{Quantum batteries represent one of the most promising applications of quantum thermodynamics, whose goal is not only to store energy inside small quantum systems but also to potentially leverage genuine quantum effects to outperform classical counterparts. In this context, however, energy fluctuations become extremely relevant and have a significant impact on the charging efficiency. In our work, we consider a simple yet paradigmatic model in which a flying qubit (the battery) coherently interacts with a single mode optical cavity (the charger) through a number conserving Jaynes-Cummings interaction.
By making use of full-counting statistics techniques, we fully characterize the average charging power, its fluctuations and the associated charging efficiency for several different choices of initial states of the optical cavity, demonstrating that preparing the latter in a genuinely quantum non-Gaussian Fock state (rather than a classical or even non-classical Gaussian state) leads to a definite and (in principle) measurable advantage in all these figures of merit.}



\end{abstract}

\maketitle

\section{Introduction}
\label{sec: Introduction}

{Since its original conception, Thermodynamics has provided principles to understand physical processes and offered optimization guidelines for the design and operation of machines, thus driving the technological progress forward. 
Such progress, over the course of the past century, has been characterized by a constant miniaturization, which has eventually led us to the construction of the first genuinely quantum devices, such as high-precision sensors~\cite{pirandola2018advances,zaiser2016enhancing,degen2017quantum}, quantum computers~\cite{Eisert1999,arute2019quantum,ladd2010quantum}, and nanoscale thermal machines~\cite{pekola2015towards,rossnagel2016single,maslennikov2019quantum,von2019spin,klaers2017squeezed,Onishchenko2024}. 
However, all of the prospective applications of these cutting-edge technologies crucially require both scalability and, even more fundamentally, an exquisite degree of accuracy in order to deliver what they promise, i.e., to outperform any classical counterpart.
Such an accuracy is quantitatively measured by the amount of fluctuations in any measured quantity of interest, which at the nanoscale may become as preponderant as their respective average values.
It is evident that having a more pronounced average output signal makes little sense or utility if its reliability drops even more significantly (i.e., its fluctuations, or its noise, grow too much). To properly talk about 'quantum advantage', it is thus necessary to take these fluctuations into account.}

Quantum stochastic thermodynamics has recently proved once again to be an invaluable guide in this respect, by demonstrating that curbing down fluctuations inevitably implies a minimum amount of energetic resources that must be irreversibly spent.
Crucially, the contribution of genuinely quantum effects, such as quantum friction or entanglement, has been singled out in the precision of thermodynamically relevant operations, e.g., the operation of heat engines.
This begs for the following central question: \textit{is it possible to achieve proper ``quantum advantage'', as specified above, in a thermodynamically meaningful task?}

Due to the inevitably much larger energy scales needed to operate and preserve quantum resources, we believe that two such tasks exist: quantum refrigeration \cite{mitchison2019refrigerators}, in which the goal is to cool a quantum system to the lowest possible temperature, and quantum battery charging~\cite{campaioli2024colloquium}, whose task is to inject, preserve, and extract energy from a quantum system. In the present work, we focus on the latter.


A quantum battery is a generic quantum system capable of storing energy after being charged, and then delivering it on demand when needed \cite{alicki2013,campaioli2024colloquium}. A vast theoretical literature has become available on this topic, mostly concerning theoretical proposals~\cite{andolina2018charger,
CampaioliPRL2017,Barra2019,caravelli2020random, Tacchino2020, shaghaghi2023lossy, yang2024three, delmonte2021characterization, Barra1, Barra2, uwefisher_beneficial}, followed only recently by the first experimental implementations \cite{hu2022optimal,joshi2022experimental, quach2022superabsorption,maillette2023experimental}.
Nevertheless, the applications and the role of quantum batteries in future technology with the aim, e.g., of enhancing the capacity of energy storage devices \cite{ferraro2019supercapacitors}, or becoming an integrated part of quantum computing processes \cite{chiribella2021reversibleQC}, still need to be thoroughly investigated. On the other hand, the interest in quantum batteries is also motivated by the usefulness that such theoretical models represent in fundamental physics studies, e.g., in understanding the role of energy fluctuations in quantum systems by using thermodynamic-based approaches. While different routes can be taken (such as the investigation of upper bounds on non-equilibrium fluctuation of observables in quantum dynamics \cite{Hamazaki_2024_speed_limit,axioms13120817_fromUncertaintyRelations}), quantum batteries are a promising test field for quantum thermodynamics itself.

\begin{figure}[ht]
    \centering
    \includegraphics[width=0.48\textwidth]{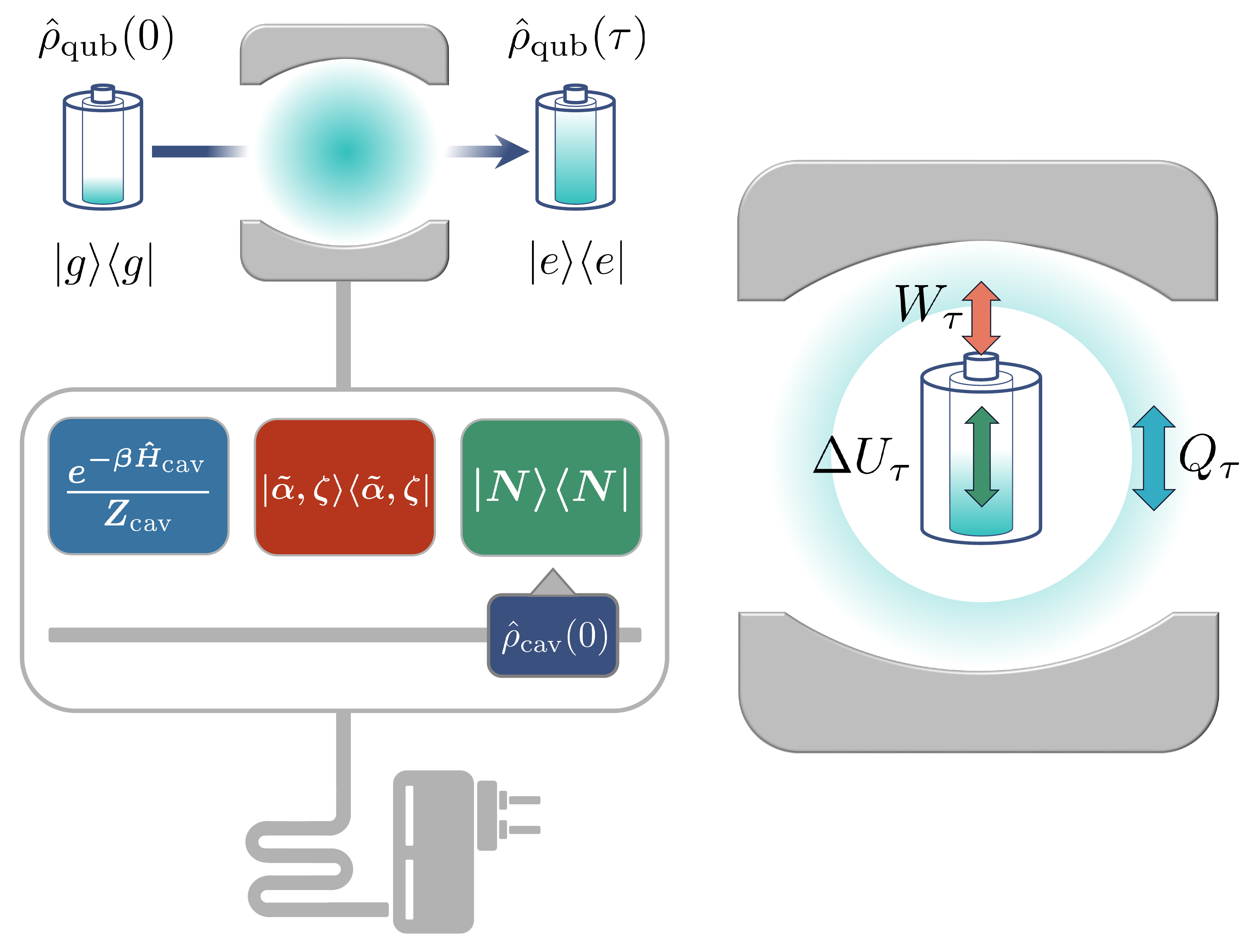}
    \caption{Schematic representation of the proposed JC quantum battery. The single-mode resonator (i.e., the charger) can be initially set in an arbitrary radiation state, $\hat{\rho}_{\text{cav}}(0)$, such as, e.g., a thermal state with inverse temperature $\beta$, a Fock state with a fixed number of photons $N$, a coherent state with an average number of photons $|\alpha|^2 = N$, or a squeezed coherent state with squeezing parameter $\zeta$ and displacing parameter $\tilde{\alpha}$ with the mean number of photons $\langle n\rangle_{\xi,\tilde{\alpha}}=N$. The flying qubit (i.e., the quantum battery), enters the cavity at time $t=0$ in its  ground state $\hat{\rho}_{\text{qub}}(0) = \ket{g}\bra{g}$. The radiation field is the resource to be used to charge the qubit. After the charging time $t=\tau>0$, the qubit state should be as close as possible to its excited state, i.e., $\hat{\rho}_{\text{qub}}(\tau) \to \ket{e}\bra{e}$. We define the relevant quantities associated with the energetic performance of the quantum battery: the moments (i.e., the average values) and the cumulants (i.e., the variances) of the internal energy variation, $\Delta U_\tau$, the heat exchanged, $Q_\tau$, and the work performed, $W_\tau$, as defined in the text.
    }
    \label{fig: FIG1}
\end{figure}

In particular, since by definition quantum batteries are subject to highly non-equilibrium processes (i.e., charging and discharging), the key quantities to be controlled are the energy fluctuations in time, i.e., the variance of its stochastic change over a given time window. In order to access these higher moments, we operate within the so-called \textit{two-point measurement scheme} (TPM), in which the well-established Full Counting Statistics (FCS) formalism can be exploited ~\cite{esposito2009nonequilibrium, brenes2023particle}. Within this framework, all the cumulants of the stochastic variable $\Delta E$, associated with the change in energy over a certain fixed time, can be evaluated.

In this work we study a closed bipartite system based on the number-conserving and coherent Jaynes-Cummings (JC) model: a flying qubit interacting with a single-mode optical cavity for a finite time, $\tau$. We call this device a \textit{JC quantum battery}, as it is represented schematically in Fig.~\ref{fig: FIG1}. The interaction process is meant to transfer energy from the radiation field (or another single-mode oscillator) populating the cavity mode in a given state (i.e., the \textit{charger}) onto the qubit (i.e., the quantum \textit{battery}), thus defining an actual charging process to all intents and purposes. For our purposes, we suppose that the qubit can be set in its ground state at the beginning of the process. On the other hand, we assume to have a high degree of control on the cavity: consequently, the confined electromagnetic field mode can be initially prepared in a specific state, such as classical thermal or coherent states, non-classical squeezed displaced \cite{adesso2014continuous} and quantum non-Gaussian Fock states. Our results show that, by preparing the cavity in a genuine quantum state (i.e., a Fock state), it is possible to achieve a significant quantum advantage in the quantum battery charging process, as compared to charging protocols that involve a charger prepared in a classical state or even non-classical Gaussian state with the same mean energy (photon number). Importantly, such an advantage is furthermore reliable and robust, since it also takes into account the power fluctuations associated with the process. 
\\
Before proceeding, let us notice that  models of a JC quantum battery have already been addressed in the literature. Among the different characterizations, an analysis similar to ours has been reported in Ref.~\cite{shaghaghi2023lossy}, although with the main purpose of charging an oscillator (a micromaser quantum battery) via a stream of qubits, each one prepared in their excited state. Similarly, three-level systems in place of qubits were employed in Ref.~\cite{Barra2}, where the possibility of swapping the role of battery and charger was also investigated. Sticking to the single-qubit JC model, a two-photon JC quantum battery was investigated in Ref.~\cite{delmonte2021characterization}, where it has been shown that the use f a Fock state allows to inject a larger amount of energy as compared to the classical counterparts (coherent and squeezed states), in line with the main results obtained in the present work. We also emphasize that the advantage of using a Fock state might depend on the physical situation that has been taken into account. As an example, another recent work analyzed the charge of multiple three-level systems coupled to a cavity mode prepared in Fock, coherent, or squeezed coherent states, respectively~\cite{yang2024three}. In the latter results, the coherent state brings some advantage when compared to the others. However, the advantage provided by the coherent state is likely due to collective effects, such as the unwanted generation of entanglement between the systems, which seems to increase when using Fock or squeezed states, thus preventing the possibility for these states to enhance the battery performances. On the contrary, our model focuses on the interaction between the cavity and a single qubit at a time, which does not allow the emergence of quantum correlations between subsequent qubits flying into the cavity region. In such a context, Fock states seem to bring a quantitative advantage over the others. \\

The paper is structured as follows: in Sec.~\ref{sec: Model} we present the model of a JC quantum battery; in Sec.~\ref{sec: Dynamical fluctuations and FCS} we introduce the fundamental thermodynamic quantities involved in this process, as well as the FCS technique used to characterize their higher fluctuations; Sec.~\ref{sec: Results} will then be devoted to the presentation of all the results, which will then be further discussed in detail in Sec.~\ref{sec: Discussion}. Section~\ref{sec: Conclusions} finally pertains to conclusions and outlooks stemming from our work.

\section{The Model}
\label{sec: Model}

We consider the case of a qubit with internal transition frequency $\omega_{\text{qub}}$, interacting for a finite time $\tau$ with a single mode of an electromagnetic field confined to a resonator, whose harmonic oscillating frequency is denoted as $\omega_{\text{cav}}$. The interaction between the qubit and the cavity mode occurs, in general, through a time-dependent coupling strength $g(t)$, which in its simplest version can be set to be nonzero (and constant within the time of flight into the resonator) only between time $t=0$ and time $t=\tau$, i.e., explicitly we have $g(t) = g\cdot\Theta(t)\Theta(\tau-t)$, where $\Theta(t)$ is the Heaviside step function: $\Theta(t) = 1$ for $t\geq0$ and $\Theta(t) = 0$ elsewhere. With this approximation in mind, therefore, we can think of the coupling strength as a constant parameter for times $0 \leq t \leq \tau$, which leads to a theoretical description in terms of the Rabi Hamiltonian \cite{braak2011integrability}, in general. However, if we restrict the analysis to the regime in which $g/\omega_0 \ll 1$, with $\omega_{\text{qub}} \simeq \omega_0 \simeq \omega_{\text{cav}}$, then the counter-rotating terms in the Rabi Hamiltonian can be neglected (the so-called rotating wave approximation \cite{stenholm1973quantum}), leaving us with the simple JC Hamiltonian \cite{stenholm1973quantum, Wallraff2004,Hennessy2007,DeLiberato2009,schuster2007resolving,forn2019ultrastrong, beaudoin2011dissipation}:
\begin{equation}
    \label{eq: JC Hamiltonian}
    \begin{split}
        \hat{H}_{\text{JC}}(t) &= \hat{H}_{\text{qub}} + \hat{H}_{\text{cav}} + \hat{H}_{\text{int}}(t) \\
    & = \hbar \omega_{\text{qub}} \frac{1}{2}\hat{\sigma}_z + \hbar \omega_{\text{cav}} \hat{a}^\dagger\hat{a} +  \hbar g(t) (\hat{\sigma}_+\hat{a} + \hat{\sigma}_-\hat{a}^\dagger)
    \end{split}
\end{equation}
being $\hat{\sigma}_z$ the $z-$Pauli matrix, $\hat{\sigma}_- (\hat{\sigma}_+$) the annihilation (creation) operator of a single qubit excitation, and $\hat{a} (\hat{a}^\dagger$) the annihilation (creation) operator of a single excitation in the cavity field mode. If we write the qubit state with respect to the basis $\{\ket{e}, \ket{g}\}$ (i.e., the one consisting of the pure excited state $\ket{e}$ and of the pure ground state $\ket{g}$), then $\hat{\sigma}_z = \ket{e}\bra{e} - \ket{g}\bra{g}$, $\hat{\sigma}_- = \ket{g}\bra{e}$, and $\hat{\sigma}_+ = \ket{e}\bra{g}$.

In general, we denote the qubit state as $\hat{\rho}_{\text{qub}}(t)$, and the cavity mode state as $\hat{\rho}_{\text{cav}}(t)$. The joint system is then defined by the state $\hat{\rho}_{\text{JC}}(t)$, and we assume that, at the initial time $t=0$, the states of the two subsystems are factorized: $\hat{\rho}_{\text{JC}}(0)=\hat{\rho}_{\text{qub}}(0)\otimes\hat{\rho}_{\text{cav}}(0)$, meaning that no correlation (whether classical or quantum) is present before the interaction. No other prescription is needed for what follows. However, since we are dealing with a charging protocol, from Sec.~\ref{sec: Results} onwards we will assume that the qubit is prepared in its ground state at time $t=0$: $\hat{\rho}_{\text{qub}}(0) = \ket{g}\bra{g}$. Likewise, we will suppose that the single-mode cavity mode can be initialized in one of the following:
\begin{equation}
    \label{eq: Cavity's states}
    \begin{split}
        &\hat{\rho}_{\text{thermal}} = \frac{e^{-\beta \hat{H}_{\text{cav}} }}{Z_{\text{cav}}} \\
    &\hat{\rho}_{\text{coherent}} = \ket{\alpha}\bra{\alpha} \\
    &\hat{\rho}_{\text{squeezed}} = \ket{\tilde{\alpha},\zeta}\bra{\tilde{\alpha},\zeta} = \hat{S}(\zeta)\ket{\tilde{\alpha}}\bra{\tilde{\alpha}}\hat{S}^\dagger(\zeta)  \\
    &\hat{\rho}_{\text{Fock}} = \ket{N}\bra{N}  \\
    \end{split}
\end{equation}
in which $Z_{\text{cav}} = \text{Tr} \{ e^{-\beta \hat{H}_{\text{cav}}} \}$ is the canonical partition function associated with the thermal state at inverse temperature $\beta$, $\ket{\alpha} = e^{-\frac{|\alpha|^2}{2}}\sum_{k=0}^{\infty}\frac{\alpha^k}{\sqrt{k!}} \ket{k}$ is the pure coherent state with $|\alpha|^2$ photons on average, $\hat{S}(\zeta) = e^{\frac{1}{2}(\zeta^\ast \hat{a}^2 - \zeta \hat{a}^{\dagger2})}$ is the squeezing operator (with squeezing parameter $\zeta$), and $N$ is the number of photons in the Fock state $\ket{N}$. We notice that the JC model has been shown to be analytically solvable \cite{stenholm1973quantum, smirne2010nakajima, bocanegra2024invariant}. Let us define the evolution operator $\hat{\mathcal{U}}(t,0) = \mathcal{T}\text{exp}\{-\frac{i}{\hbar}\int_{0}^{t}\hat{H}_{\text{JC}}(t')dt'\}$ (being $\mathcal{T}$ the time-ordering operator) such that $\hat{\rho}_{\text{JC}}(t) = \hat{\mathcal{U}}(t,0) \hat{\rho}_{\text{JC}}(0) \hat{\mathcal{U}}^\dagger(t,0)$: such operator has a well known expression, and it can be explicitly written as a $2\times2$ matrix of cavity operators. For completeness, we provide it in App.~\ref{app: JC evolution operator}, see Eq.~(\ref{eq: Evolution operator of JC, analytical}).

\section{Dynamical fluctuations and Full Counting Statistics}
\label{sec: Dynamical fluctuations and FCS}
Since we are interested in the energy exchanges between two quantum systems during a time window $\tau$, and most of all in their fluctuations, we must go beyond the \textit{static} description of the involved meaningful quantities, i.e., the one given by the moments of quantum observables $\hat{A}(t)$ at a fixed time $t$, such as $\langle \hat{A}^n(t) \rangle = \text{Tr}\{ \hat{A}^n(t)\hat{\rho}(t)\}$. These are \textit{state-dependent} objects, and do not capture the essence of the non-equilibrium physics behind the battery charge. On the contrary, we shall look at the \textit{dynamical} observables \cite{hickey2013time}, i.e., stochastic variables $\Delta a_\tau$ that can be related to the measurement outcomes of the corresponding observables $\hat{A}(t)$ at two different times. These are, by definition, \textit{process-dependent} quantities, and allow for a proper investigation of the out-of-equilibrium phenomena. The paradigmatic way to connect $\Delta a_\tau$ with $\hat{A}(t)$ is by the TPM scheme \cite{esposito2009nonequilibrium, gherardini2024quasiprobabilities}, in which $\Delta a_\tau \equiv a_\tau - a_0$ is defined as the difference between two outcomes, $a_\tau$ and $a_0$, obtained by the projective measurement of the observable $\hat{A}(t)$ at time $t=\tau$ and at time $t=0$, respectively. 

As an example, if we identify $\hat{A} = \hat{H}_{\text{qub}}$ (i.e., the qubit Hamiltonian), then $\Delta a_\tau \rightarrow \Delta U_\tau$ is a stochastic variable that represents the qubit energy variation $\Delta U$ over the time window $\tau$, associated with two sequential energy measurements performed on the qubit at different times. As the notations suggests, we will denote the qubit energy variation with $\Delta U_\tau$, that is the \textit{internal energy variation} of the system of interest (i.e., the battery). Besides, we will consider as \textit{heat} the change in energy of the environment~\cite{landi2021irreversible}: in this case, it is the energy change of the cavity (i.e., the charger), which actually acts as a sort of environment. Therefore, by identifying $\hat{A}$ with $\hat{H}_{\text{cav}}$, the heat is $\Delta a_\tau \rightarrow Q_\tau$, which is linked to two sequential energy measurements made on the cavity. Finally, by employing the first law of thermodynamics, we get the \textit{work} $W_\tau$ as the energy contribution stemming from the energy conservation, i.e., $\Delta U_\tau = Q_\tau + W_\tau$. This contribution can be interpreted as the amount of energy stored in the correlations between the two systems, in analogy with the expectation value of the interaction Hamiltonian at time $t=\tau$, $\langle \hat{H}_{\text{int}}(\tau)\rangle$. Anyhow, the latter is no more than an analogy: as discussed before, $\langle W_\tau\rangle$ is the average value of a process-dependent stochastic variable over a time window $\tau$, while $\langle \hat{H}_{\text{int}}(\tau)\rangle$ is a state-dependent quantum mechanical expectation value at time $t=\tau$. Although they are quantitatively identical at the level of average values, their interpretation is conceptually different, indeed. We notice, for example, that the variance of $W_\tau$ does not coincide with the state-dependent variance $\langle \hat{H}^2_{\text{int}}(\tau)\rangle - \langle \hat{H}_{\text{int}}(\tau)\rangle^2$, in general. \\
As it will be shown in Eqs. (\ref{eq: SIMPLE RELATIONS}), $W_\tau$ vanishes in case of perfect resonance (i.e., whenever the JC model is energy conserving), while it is nonzero if $\Delta\omega \neq 0$. \\


The next step is to make use of FCS in order to compute the statistical cumulants of such quantities. A complete treatment on FCS is provided in Ref.~\cite{esposito2009nonequilibrium}: here we briefly summarize only the key concepts. In particular, in order to find the statistics of $\Delta U_\tau$, $Q_\tau$, and $W_\tau$, we need to find a generating function $\mathscr{G}_\tau(\boldsymbol{\chi})$ depending on the so-called \textit{counting parameters}, $\chi_1$ and $\chi_2$, summarized by the vector $\boldsymbol{\chi}=(\chi_1, \chi_2)$. The counting parameter $\chi_1$ is associated with the qubit change in energy, i.e., $\Delta U_\tau$, while $\chi_2$ is associated with the variation of energy stored in the cavity mode, that is $Q_\tau$. This relation is formalized by defining the \textit{tilted} evolution operator
\begin{equation}
    \label{eq: Tilted evolution operator}
        \hat{\mathcal{U}}_{\frac{\boldsymbol{\chi}}{2}}(\tau,0) = e^{i\frac{\chi_1}{2}\hat{H}_{\text{qub}} - i\frac{\chi_2}{2}\hat{H}_{\text{cav}}}\hat{\mathcal{U}}(\tau,0)e^{-i\frac{\chi_1}{2}\hat{H}_{\text{qub}} + i\frac{\chi_2}{2}\hat{H}_{\text{cav}}}
\end{equation}
obtained from the known evolution operator $\hat{\mathcal{U}}(\tau,0)$ of the JC model. As shown in Eq.~(\ref{eq: Tilted evolution operator}), we notice that $\chi_1$ multiplies $\hat{H}_{\text{qub}}$, while $\chi_2$ multiplies $\hat{H}_{\text{cav}}$, and their sign is conventionally chosen, without any loss of generality, in order to match the sign convention of the First Law of Thermodynamics. It turns out that the generating function can be found simply by taking the trace of the joint systems density matrix at the starting time, $\hat{\rho}_{\text{JC}}(0)$, multiplied by the tilted evolution operator $\hat{\mathcal{U}}_{\frac{\boldsymbol{\chi}}{2}}(\tau,0)$ from the left, and by $\hat{\mathcal{U}}^\dagger_{-\frac{\boldsymbol{\chi}}{2}}(\tau,0)$ from the right:
\begin{equation}
    \label{eq: Generating function}
    \mathcal{G}_\tau(\boldsymbol{\chi}) = \text{Tr}\bigl\{ \hat{\mathcal{U}}_{\frac{\boldsymbol{\chi}}{2}}(\tau,0) \hat{\rho}_{\text{JC}}(0) \hat{\mathcal{U}}^\dagger_{-\frac{\boldsymbol{\chi}}{2}}(\tau,0)\bigr\}
\end{equation}
Once the generating function is known, it is possible to find the statistical moments of $\Delta U_\tau$ and $Q_\tau$, by performing the derivative of $\mathscr{G}_\tau(\boldsymbol{\chi})$ with respect to $\chi_1$ and $\chi_2$, respectively, and by setting $\chi_1 = 0 = \chi_2$. The statistics of the work $W_\tau$ can then be obtained by using the first law of thermodynamics:
\begin{equation}
    \label{eq: U, Q, W}
    \begin{split}
\langle\Delta U_\tau^n\rangle &= (-i)^n \frac{\partial^n\mathcal{G}(\boldsymbol{\chi})}{\partial\chi_1^n}\biggl\rvert_{\chi_1 = 0 = \chi_2} \\
\langle Q_\tau^n\rangle &= (-i)^n \frac{\partial^n\mathcal{G}(\boldsymbol{\chi})}{\partial\chi_2^n}\biggl\rvert_{\chi_1 = 0 = \chi_2} \\
\langle W_\tau^n \rangle &= \langle(\Delta U_\tau - Q_\tau)^n \rangle
    \end{split}
\end{equation}
Finally, the cumulants of the stochastic variable $Z$, where $Z \in \{\Delta U_\tau, Q_\tau, W_\tau\}$, can be found by employing the definition $\kappa_n(Z) = \langle (Z - \langle Z \rangle)^n \rangle$. As an example, the second-order cumulant of $Z$ is its variance, i.e., $\text{var}(Z) = \langle (Z - \langle Z \rangle)^2 \rangle = \langle Z^2 \rangle - \langle Z \rangle^2$. 
The detailed expressions for $\hat{\mathcal{U}}_{\frac{\boldsymbol{\chi}}{2}}$ and $\mathscr{G}_\tau(\boldsymbol{\chi})$ are provided in App.~\ref{app: the generating function}, Eqs.~(\ref{eq: Tilted U(Chi/2), explicit (appendix)}) and (\ref{eq: Generating function GENERAL}).

\section{Results}
\label{sec: Results}
Here we present the main result of the present work, namely, the possibility of exploiting a non-classical cavity initial state in order to achieve an effective quantum advantage in the charging process of a JC quantum battery, even at the level of energy fluctuations. In particular, the signal-to-noise ratio (SNR) associated with a charging protocol where the cavity is prepared in a Fock state, $\ket{N}\bra{N}$, reaches much higher values than any other protocol based on Gaussian (i.e., thermal, coherent, or squeezed coherent) states of light sharing the same average energy. This result is concisely shown in Fig.~(\ref{fig: FIG2}), while the steps necessary to obtain it and a thorough discussion are given below.

\subsection{The signal-to-noise ratio}
\label{subsec: The SNR}
First, we define the SNR associated with a stochastic observable $Z$:
\begin{equation}
    \label{eq: SNR definition}
    \text{SNR}(Z) = \frac{\langle Z \rangle^2}{\text{var}(Z)} 
\end{equation}
which quantifies the relative quality of the \textit{signal} (i.e., the squared average $\langle Z \rangle^2$) with respect to its \textit{fluctuations} (that is, the variance $\text{var}(Z)$). In our discussion, $Z$ denotes a stochastic quantity between $\Delta U_\tau$, $Q_\tau$, and $W_\tau$.

\begin{figure}[t]
    \centering
    \includegraphics[width=\linewidth]{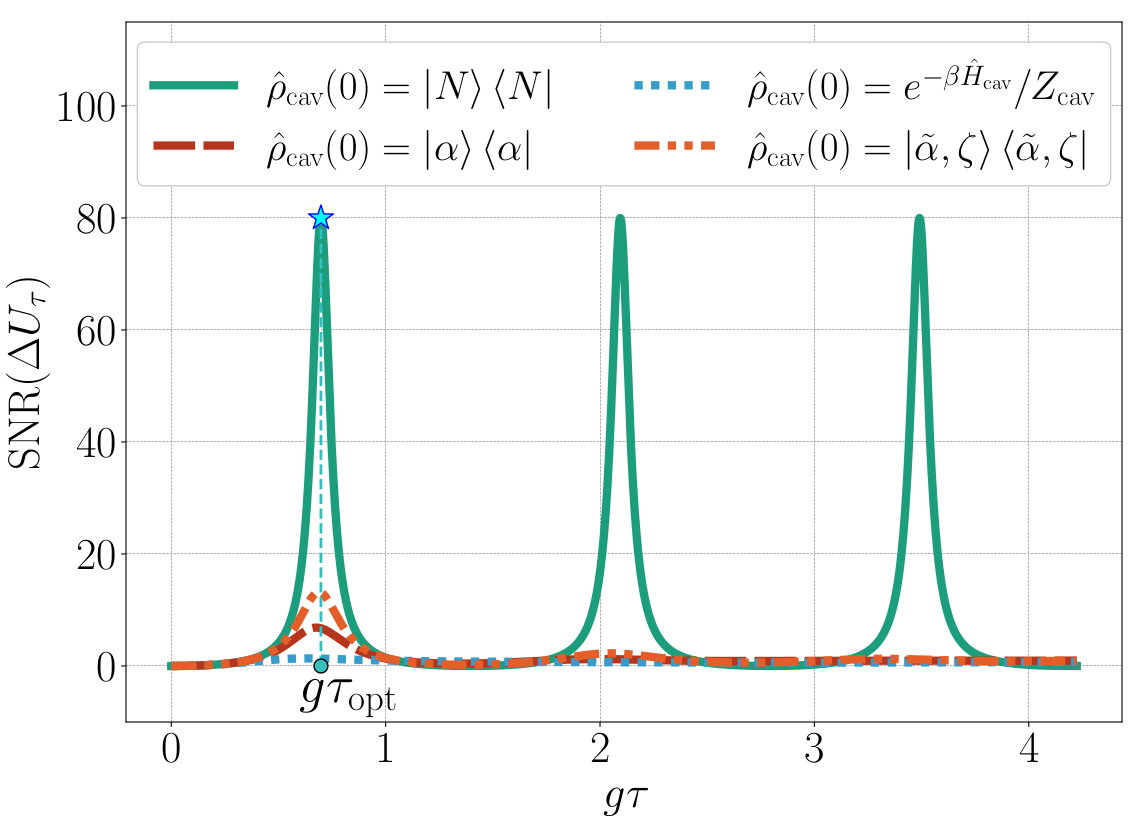}
    \caption{ Plot of the $\text{SNR}(\Delta U_\tau)$ from Eq.~(\ref{eq: SNR definition}) as a function of the time window $g\tau$, related to four different initial cavity states: pure Fock (continuous green line), coherent (dashed red), thermal (dotted blue), and squeezed coherent (dash-dotted orange). Importantly for a fair comparison between them, all these states have the same average photon number, which is $\langle n \rangle = N = 5$. Moreover, here we set the other parameters to be $\omega_{\text{qub}} = 1 \text{ GHz}$, $g/\omega_{\text{qub}} = 10^{-2}$, $\Delta\omega = \omega_{\text{qub}} - \omega_{\text{cav}}$ and $\Delta\omega/\omega_{\text{qub}} = 5\cdot10^{-3}$, $\beta = \log(\frac{N+1}{N}) / \hbar\omega_{\text{cav}}$, $\alpha = \sqrt{N}$, $\zeta = 0.6$, and $\tilde{\alpha} \simeq 3.906$ (in order to maintain the constraint $\langle n \rangle = N = 5$). As reported in Eq.~(\ref{eq: SNR equivalence}), $\text{SNR}(\Delta U_\tau)$, $\text{SNR}(Q_\tau)$, and $\text{SNR}(W_\tau)$ are quantitatively identical.}
    \label{fig: FIG2}
\end{figure}

As was mentioned before, although the analysis can be carried out without imposing any assumption on the initial states of the qubit and the cavity mode (i.e., the framework presented in Sec.~\ref{sec: Dynamical fluctuations and FCS} is completely general), we now restrict ourselves to the particular case of a qubit initial state $\hat{\rho}_{\text{qub}}(0)=\ket{g}\bra{g}$, and a cavity mode initially in a Fock state $\hat{\rho}_{\text{cav}}(0)=\ket{N}\bra{N}$. This means that the battery is, in principle, discharged, while the charger is in a quantum state corresponding to exactly $N$ photons stored. Also, we focus only on the computation of the statistics related to the variation of the qubit energy, i.e. $\Delta U_\tau$: for that reason, we set $\chi_2 = 0$ in the generating function, in order to discard the factors associated with $Q_\tau$, and to further enlighten the following expressions. The details regarding the derivation are provided in App.~\ref{app: the generating function}.

Therefore, under these assumptions, the generating function $\mathcal{G}_\tau(\chi_1,\chi_2=0) \equiv \mathcal{G}_\tau(\chi_1)$ has the following form:
\begin{equation}
    \label{eq: Generating function, particular case}
    \mathcal{G}_\tau(\chi_1) = g^2 N e^{i\chi_1\hbar\omega_{\text{qub}}} \mathscr{S}^2(\tau) + \mathscr{C}^2(\tau) + \bigl(\frac{\Delta\omega}{2}\bigr)^2 \mathscr{S}^2(\tau)
\end{equation}
where the functions $\mathscr{S}(\tau)$ and  $\mathscr{C}(\tau)$ are shortcuts for:
\begin{equation}
      \label{eq: S(t) and C(t) functions}
\begin{split}
    \mathscr{S}(\tau) &= \frac{\sin \biggl( \tau\sqrt{g^2 N + (\frac{\Delta\omega}{2})^2 } \biggr) }{\sqrt{g^2 N + (\frac{\Delta\omega}{2})^2 }} \\ 
    \mathscr{C}(\tau) &= \cos \biggl( \tau\sqrt{g^2 N + (\frac{\Delta\omega}{2})^2 } \biggr)
\end{split}
\end{equation}
and $\Delta\omega = \omega_{\text{qub}} - \omega_{\text{cav}}$ is the detuning between the qubit and the cavity. By taking the first derivative with respect to $\chi_1$ as in Eq.s~(\ref{eq: U, Q, W}), and by setting $\chi_1=0$, we then obtain the analytic expression for the change in energy of the qubit:
\begin{equation}
    \label{eq: 1st momentum U}
    \langle \Delta U_\tau \rangle = g^2 N \hbar \omega_{\text{qub}} \mathscr{S}^2(\tau) 
\end{equation}
It is then straightforward to compute the second-order momentum of $\Delta U_\tau$, i.e., $\langle \Delta U_\tau^2 \rangle$, from which we get the variance $\text{var}(\Delta U_\tau) = \langle\Delta U_\tau^2 \rangle - \langle\Delta U_\tau \rangle^2$:
\begin{equation}
    \label{eq: variance U}
    \begin{split}
    \text{var}(\Delta U_\tau) &= g^2 N (\hbar \omega_{\text{qub}})^2 \mathscr{S}^2(\tau) \\
    &\quad - [g^2 N \hbar \omega_{\text{qub}} \mathscr{S}^2(\tau)]^2      
    \end{split}
\end{equation}
Finally, the SNR can be found from Eqs.~(\ref{eq: 1st momentum U}) and (\ref{eq: variance U}), by dividing the square of $\Delta U_\tau$ by $\text{var}(\Delta U_\tau)$:
\begin{equation}
    \label{eq: SNR Fock}
    \text{SNR}(\Delta U_\tau) = \frac{g^2 N \mathscr{S}^2(\tau) }{1 - g^2 N \mathscr{S}^2(\tau)} 
\end{equation}
Analogously, we can calculate the SNR for different initial cavity states, e.g., the ones defined in Eq.~(\ref{eq: Cavity's states}). Although an analytical formulation is still possible, in these cases it involves summations over infinite terms (i.e., over the cavity energy levels): therefore, computations must be carried out numerically, by truncating the summations. The results corresponding to four different initial cavity mode states are reported in Fig.~\ref{fig: FIG2}. Here we have fixed $N = 5$, and consequently we choose $\beta = \log(\frac{N+1}{N}) / \hbar\omega_{\text{cav}}$ (by direct application of the Bose-Einstein statistical relation), and $\alpha = \sqrt{N}$; regarding the squeezed coherent state, we choose a real squeezing parameter $\zeta = 0.6$, and we picked $\tilde{\alpha} \simeq 3.906$. In this way, each Gaussian state contains $\langle n \rangle = N = 5$ photons on average. Therefore, a direct comparison can be made between the different cases, showing that an effective quantum advantage in the charging process of a JC quantum battery is achieved when the cavity mode is prepared in a Fock state as compared to any Gaussian (i.e., thermal, coherent, or squeezed coherent) states of light, as anticipated at the beginning of this Section.

Additionally, it is possible to demonstrate that the moments and the cumulants of the stochastic quantities $Q_\tau$ and $W_\tau$ are directly proportional to the corresponding moments and cumulants of the internal energy variation, $\Delta U_\tau$, with proportionality coefficients that depend only on the characteristic frequencies of the qubit and the cavity mode, i.e.,
\begin{equation}
    \label{eq: SIMPLE RELATIONS}
    \begin{split}
    \langle Q^n_\tau \rangle &= \biggl( \frac{\omega_{\text{cav}}}{\omega_{\text{qub}}} \biggr)^n\langle \Delta U_\tau^n \rangle \\
    \kappa_n(Q_\tau) &= \biggl( \frac{\omega_{\text{cav}}}{\omega_{\text{qub}}} \biggr)^n \kappa_n(\Delta U_\tau) \\
    \langle W_\tau^n \rangle &= \biggl( 1 -\frac{\omega_{\text{cav}}}{\omega_{\text{qub}}} \biggr)^n \langle \Delta U_\tau^n \rangle \\
    \kappa_n(W_\tau) &= \biggl( 1 -  \frac{\omega_{\text{cav}}}{\omega_{\text{qub}}} \biggr)^n \kappa_n(\Delta U_\tau)
    \end{split}
\end{equation}
The derivation of these relations is detailed in App.~\ref{app: Sketch of proof}. It is interesting to highlight that, albeit simple, they are completely general in the sense that they do not depend on any other physical parameter but the ratio between $\omega_{\text{cav}}$ and $\omega_{\text{qub}}$ (for example, they do not even depend on the coupling strength $g$, or the average number of photons in the cavity). Besides, they neither depend on the initial system states, thus remaining valid even beyond the assumptions under which we derived Eq.~(\ref{eq: SNR Fock}). Furthermore, they hold for moments and cumulants of any order $n$.

Such a simple analytic form is ultimately due to the relative simplicity of the JC model itself, together with the generality of the first law of Thermodynamics, for which, given the two energy contributions $\Delta U_\tau$ and $Q_\tau$, the third one, $W_\tau$, is automatically quantified by the energy conservation argument. The simplicity of these relations is reflected in the SNR: in fact, once the initial states of qubit and cavity have been fixed, it is sufficient to compute the average and the variance of $\Delta U_\tau$ to obtain the ones of $Q_\tau$ and $W_\tau$, by scaling $\langle\Delta U_\tau\rangle$ and $\text{var}(\Delta U_\tau)$ with the factors $\omega_{\text{cav}}/\omega_{\text{qub}}$ and $(\omega_{\text{cav}}/\omega_{\text{qub}})^2$, respectively. Therefore, we have that
\begin{equation}
    \label{eq: SNR equivalence}
    \text{SNR}(\Delta U_\tau) = \text{SNR}(Q_\tau) = \text{SNR}(W_\tau)
\end{equation} because the ratio $(\omega_{\text{cav}}/\omega_{\text{qub}})^2$ simplifies out. This is the reason why, in Fig.~\ref{fig: FIG2}, the SNR has been computed from the expectation and the variance of $\Delta U_\tau$ only.

\subsection{Charging protocol: power and charge efficacy}
\label{subsec: Charging protocol, power and charge efficacy}
Following Ref.~\cite{campaioli2024colloquium}, we define the power $P_\tau$ as the quantity of energy injected in the qubit per unit of time, i.e.,
\begin{equation}
    \label{eq: Power generic}
    \langle P_\tau \rangle \equiv \frac{\langle \Delta U_\tau \rangle}{\tau}
\end{equation}
Under the same conditions that led to Eq.~(\ref{eq: SNR Fock}), i.e. assuming that the qubit and the cavity mode start in the ground state and in a Fock state, respectively, it is possible to analytically derive the optimal charging time $\tau_{\text{opt}}$, defined as the shortest time for which the SNR (\ref{eq: SNR Fock}) is maximized. 
This condition is achieved when the sine contained in the function $\mathscr{S}(\tau)$ of Eqs.~(\ref{eq: S(t) and C(t) functions}) is equal to 1:
\begin{equation}
    \label{eq: Optimal charging time}
    \tau_{\text{opt}} = \frac{\pi}{2\sqrt{g^2 N + (\frac{\Delta\omega}{2})^2}}
\end{equation}
From the above expression, we can recover the following result: $ \tau_{\text{opt}} \sim 1/\sqrt{N}$. The latter describes the speedup that can be obtained by using a cavity Fock state with $N$ photons \cite{andolina2018charger}, which is a useful cross-check. We also remark here that an analytic expression for $\tau_{\text{opt}}$ is only possible for the Fock state cavity, while for all the other initial states, a numerical solution is required. 

\begin{figure}[t]
    \centering
    \includegraphics[width=\linewidth]{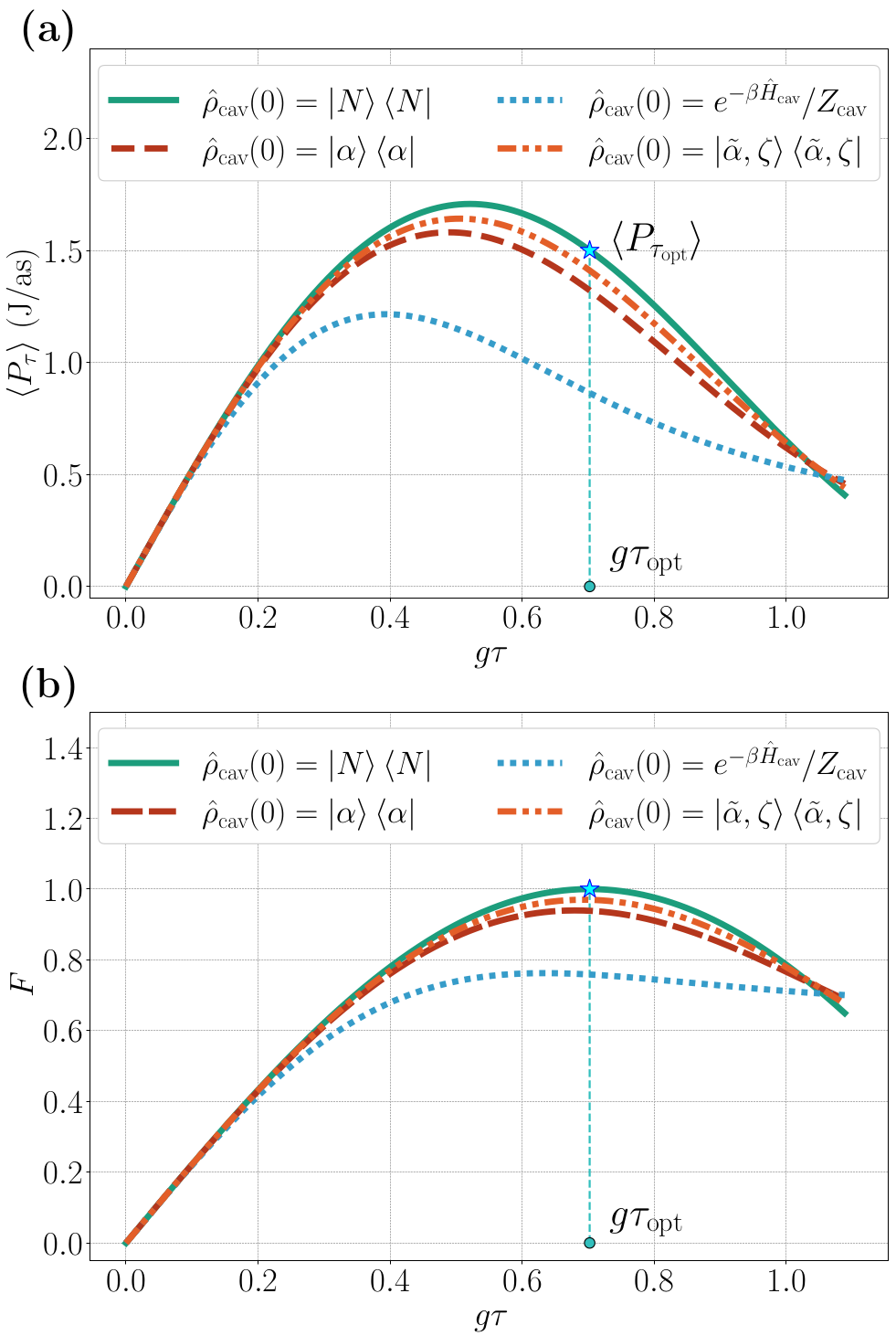}
    \caption{(a) Produced power $\langle P_\tau \rangle$, as defined in Eq.~(\ref{eq: Power generic}), plotted as a function of flying qubit time. As in Fig.~\ref{fig: FIG2}, $\langle P_\tau \rangle$ has been computed for four different initial cavity states (the color code is the same as in Fig.~\ref{fig: FIG2}). Here we assumed: $\Delta\omega/\omega_{\text{qub}} \equiv 0$. As before, we set $g/\omega_{\text{qub}} = 10^{-2}$, and we fixed the mean number of photons present in each state to be equal to $\langle n \rangle = N = 5$. (b) Fidelity $F(\hat{\rho}_{\text{qub}}(\tau), \hat{\rho}_{\text{qub}}^{\text{exc}})$ between the qubit state at time $\tau$, $\hat{\rho}_{\text{qub}}(\tau)$, and the target qubit state $\hat{\rho}_{\text{qub}}^{\text{exc}} = \ket{e}\bra{e}$, as defined in Eq.~(\ref{eq: Fidelity}). The choice of the simulation parameters is exactly the same as in Fig.\ref{fig: FIG3}a.}
    \label{fig: FIG3}
\end{figure}
By plugging Eq.~(\ref{eq: 1st momentum U}) into Eq.~(\ref{eq: Power generic}) and evaluating it over the optimal time window $\tau = \tau_{\text{opt}}$, we find: 
\begin{equation}
    \label{eq: Power at optimal tau}
    \langle P_{\tau_{\text{opt}}} \rangle = \frac{2g^2 N \hbar\omega_{\text{qub}}}{\pi\sqrt{g^2 N + (\frac{\Delta\omega}{2})^2}}
\end{equation}
which is the produced power over the time window $\tau_{\text{opt}}$, corresponding to the attainment of the maximum SNR in the case of a Fock state cavity. 
We notice that this optimal time does not correspond to the time at which the average power output is maximum, since at the latter the energy fluctuations are still higher and consequently the relative quality of the output signal (quantified by the SNR) is lower.

Finally, we can evaluate the charging efficacy by means of the evaluation of the distance between the qubit state at the generic time $\tau$, $\hat{\rho}_{\text{qub}}(\tau)$, and the qubit target state, that is the excited state $\hat{\rho}_{\text{qub}}^{\text{exc}} = \ket{e}\bra{e}$. The goal is to bring the qubit as close as possible to its excited state, in order to maximize the battery charge. Ideally, in that case the distance between the two states should vanish. In particular, we chose the fidelity $F(\hat{\rho}_1, \hat{\rho}_2)$ \cite{sommers2003bures, alsing2023comparing, nielsen_chuang} as the proper figure of merit,  defined as
\begin{equation}
    \label{eq: Fidelity}
    F(\hat{\rho}_1, \hat{\rho}_2) = \bigl(\text{Tr}\{ \sqrt{ \sqrt{\hat{\rho}_1} \hat{\rho}_2 \sqrt{\hat{\rho}_1}  } \} \bigr)^2   
\end{equation}
where we substitute $\hat{\rho}_1 = \hat{\rho}_{\text{qub}}(\tau)$ and $\hat{\rho}_2 = \hat{\rho}_{\text{qub}}^{\text{exc}} = \ket{e}\bra{e}$.

The simulated results showing the produced power and the fidelity evaluated over a time window $g\tau$ are depicted in Fig.~\ref{fig: FIG3}. First of all, we immediately notice how the Fock state cavity produces slightly more power than the three Gaussian states. (b) Fidelity $F$, as defined in Eq.~(\ref{eq: Fidelity}), between $\hat{\rho}_{\text{qub}}(\tau)$ and $\hat{\rho}_{\text{qub}}^{\text{exc}}$, where $\hat{\rho}_{\text{qub}}^{\text{exc}} = \ket{e}\bra{e}$. The optimal charge (i.e., the process for which the final qubit state approaches $\ket{e}\bra{e}$), can be obtained only when the cavity is prepared in a Fock state. In that case, $F \to 1$ when the qubit and the cavity mode are perfectly resonant, and when the charging time is the optimal time of Eq.~(\ref{eq: Optimal charging time}).
It is worth emphasising that the charging efficacy is maximized at the very same value of time $\tau_{\text{opt}}$ that optimized the SNR. This fact, far from obvious, strongly stresses the importance of incorporating the role of energy fluctuations into the picture when characterizing the performance of thermodynamic processes. Finally, since our goal is not to characterize the thermodynamics of a generic process, but rather to focus on the specific charging protocol which brings any qubit initial state to the target `charged' state $\ket{e}\bra{e}$, the injected energy $\langle\Delta U\rangle$ can be seen to correspond to the more commonly considered \textit{ergotropy} \cite{QuantumCoherenceAndErgotropy, delmonte2021characterization, yang2024three, Allahverdyan_2004} (i.e., the maximum extractable work from a system via a unitary operation).



\section{Discussion}
\label{sec: Discussion}
The advantage in the charging protocol provided by employing a cavity in an initial Fock state is inferred from the SNR of Eq.~(\ref{eq: SNR Fock}) reaching larger values as compared to the same quantity computed by using Gaussian  states (the ones that can be associated with classical light, i.e., thermal, coherent, or squeezed coherent states, respectively), as shown in Fig.~\ref{fig: FIG2}. This means that not only the average energy injected in the qubit within a time $\tau$ is larger in the Fock state case (as it is evinced from the calculated average power in Fig.~\ref{fig: FIG3}a), but also that energy fluctuations during the charging process are minimized by preparing the cavity mode in a genuine quantum state. Besides, we notice that, in the specific case of Eq.~(\ref{eq: SNR Fock}) and in the ideal situation of perfectly resonant oscillators, $\Delta\omega = \omega_{\text{qub}} - \omega_{\text{cav}} = 0$, the SNR diverges for $\tau = \tau_{\text{opt}}$. This can be understood as follows: if the qubit is perfectly resonant with the single mode cavity, a coherent exchange of energy between the two systems takes place. In other words, as much energy is given off by the charger as it is absorbed by the battery. Moreover, the energy fluctuations in the process ideally tend to zero, making the energy exchange maximally precise. Remarkably, from Fig.~\ref{fig: FIG3}b we can also infer that the excited state of the qubit can be obtained \textit{only} if the cavity mode is prepared in a Fock state. In all the other cases, the qubit gains some energy with respect to the beginning of the charging protocol, but the energy enhancement at the end of the process is not maximal: indeed, even if the fidelity is maximized for certain time windows $g\tau$, it is exactly 1 only for the case involving a genuine quantum state. \\
It is worth remarking that, within the quantum batteries community, the expression ``advantage'' usually refers to a collective enhancement of the charging power (also defined the \textit{superlinear scaling} of power \cite{Ferraro_high_power_collective, Heisenberg_spin, Extended_Dicke}), which is an intrinsically many-body effect. Instead, the advantage that we investigated is primarily referred to a single-body quantum battery (a single qubit), and most of all to the efficacy of a single charging process. The multi-body advantage is currently under investigation.
\\
Finally, we would like to emphasize that our model could be employed to describe physical setups involving \textit{flying} qubits (e.g., atoms or ions flying through the confined electromagnetic field mode of an open resonator) as well as \textit{static} ones (for example, superconducting qubits in circuit QED, or  trapped charged ions with sufficiently long trapping times, or even semiconductor quantum dots in high-quality microresonators). In this respect, we believe that an implementation based on flying qubits would be more suitable for the realization of a valid quantum battery, since it would be possible to charge multiple qubits without resetting them immediately after the process. The reset would be unavoidable, instead, in the case of a static qubit: it has to be brought back to the ground state after each iteration step, in order to be recycled to perform a multiple charging process. Nevertheless, a physical setup based on static qubits could still be exploited to obtain a proof-of-principle demonstration of the proposed protocol.

\section{Conclusions}
\label{sec: Conclusions}
In summary, we have thoroughly discussed the charging protocol of a quantum battery constituted by a flying qubit (the battery) travelling across an optical cavity (the charger), by also taking into account the energy fluctuations occurring during the charging process. We used the Jaynes-Cummings (JC) model to describe the interaction between the qubit and a single-mode electromagnetic field confined into the cavity. This model has been shown to describe a variety of physical systems in which quantum batteries could be implemented, from superconducting circuits~\cite{BlaisRMP2021} to trapped ions in optical cavities and neutral atoms in microwave or optical resonators~\cite{Haroche_exploring2006}. By exploiting the exact expression of the JC evolution operator, we used the Full Counting Statistics (FCS) to exactly compute the cumulants of the energy exchanged between the qubit and the cavity. The calculation of the signal-to-noise ratio (SNR) revealed that, by employing a cavity prepared in a pure quantum state (i.e., in a Fock number state), it is possible to obtain a marked quantum advantage over classical charging protocols (i.e., processes involving a cavity initialized in a Gaussian state). The advantage is ``reliable'' in the sense that it is persistent even in the presence of energy fluctuations, which are reduced if a Fock state cavity is considered. Furthermore, the study of the fidelity between the qubit state at time $\tau$ and the target state, i.e., the excited state of the qubit, shows that the qubit can be brought into its excited state with unit fidelity only if the cavity has been prepared in a Fock state. Notwithstanding the simplicity of the model, it is possible to devise a more realistic situation by considering non-resonant systems and a non-perfect state preparation. 
Remarkably, an advantage of quantum state over Gaussian ones in terms of fidelity can still be found: most of all, it is stable under the variation of some non-ideal parameters, such as the detuning between qubit and cavity. 

Further developments of the theoretical model analysed in the present work will consider the presence of energy losses on the systems, and will take into account the counter-rotating terms that have been neglected within the rotating wave approximation. Moreover even the temporal profile of the coupling can be optimized in order to fit more realistic scenarios \cite{flying_piccione}. A topic of major interest will also be the characterization of many-body quantum batteries, in which $N$ qubits could be arranged in series or in parallel to fly within the cavity charger. As preliminary computations suggest, the advantage of using cavity modes initially prepared in Fock number states could be even more pronounced in the case of a sequential charge of the qubits. \\
In perspective, we are also investigating the possibility of implementing the quantum battery on a trapped $\text{}^{40}\text{Ca}^+$ ions platform \cite{mika2022single, podhora2022quantum, ritboon2022sequential}. There, the oscillator consists of the harmonic motion of the ion itself, while the qubit is represented by the first two electronic energy levels of the same ion. Further technical elaboration of the presented model and conceptual idea, including the introduction of imperfect preparation and detection, will be developed in order to make it feasible on that physical platform.  
\\
As a final consideration, we reckon that these findings can be particularly suitable for the thermodynamic analysis of a wide variety of quantum processes, especially the ones for which energy fluctuations play a prominent role, e.g. in quantum catalytic processes \cite{quantumcatalysis}. Moreover, they provide a contribution to recent results concerning the investigation of the energetics of bipartite systems \cite{maffei2021probing, prasad2024closing}, which fit well with the model discussed in the present work. 

\section*{Acknowledgements}

The authors are grateful to S. P. Prasad, M. Maffei, A. Auff\`eves, M. Kol\'a\v{r}, M. B. Arjmandi, L. Slodi{\v{c}}ka, G. Benenti, and L. Razzoli for several fruitful scientific discussions. D.R. acknowledges the kind hospitality of the Majulab at the Centre for Quantum Technologies (CQT) in Singapore, as well as the warm welcome of the Optics Department at the Palack\'{y} University in Olomouc. D.G. acknowledges the `National Quantum Science Technology Institute' (NQSTI, PE4, Spoke 1) within the PNRR  project PE0000023. G.G. acknowledges funding from the Italian Ministry of Research (MUR) under the ``Rita Levi Montalcini'' project THESEUS.
R.F. acknowledges the grant 23-06308S from the Czech Science Foundation, Project CZ.02.01.01/00/22-008/0004649 of the MEYS Czech Republic supported by the EU funding. R.F. also acknowledges funding from the MEYS of the Czech Republic (Grant Agreement 8C22001), and Project SPARQL funded from the EU Horizon 2020 Research and Innovation Programme under Grant Agreements No. 731473 and 101017733 (QuantERA).


%


\appendix
\section{The JC evolution operator}
\label{app: JC evolution operator}
The evolution operator of the JC model was initially derived in Ref.~\cite{stenholm1973quantum}, whose result turned out to be crucial for deriving the analytic evolution of the joint state of qubit and cavity mode~\cite{smirne2010nakajima, bocanegra2024invariant}. It was shown that this operator, formally defined $\hat{\mathcal{U}}(\tau,0)$, can be written as a $2\times2$ matrix of cavity operators, whose components, $\hat{\mathcal{U}}_{ij} = \hat{\mathcal{U}}_{ij}(\tau)$, are explicitly given as:
\begin{equation}
    \label{eq: Evolution operator of JC, analytical}
    \begin{split}
        \hat{\mathcal{U}}_{00} &=  e^{-\frac{i}{2}\omega_{\text{cav}} (\hat{\mathds{I}}+2 \hat{a}^\dagger\hat{a})\tau}\biggl[\mathcal{C}(\tau, \hat{\varphi} + g^2\hat{\mathds{I}}) -  i\frac{\Delta\omega}{2}\mathcal{S}(\tau, \hat{\varphi} + g^2\hat{\mathds{I}})\biggr] \\
        \hat{\mathcal{U}}_{01} &= -i g e^{-\frac{i}{2}\omega_{\text{cav}} \tau} e^{-i\omega_{\text{cav}} \hat{a}^\dagger\hat{a}\tau} \mathcal{S}(\tau, \hat{\varphi}+ g^2\hat{\mathds{I}}) \hat{a}  \\
        \hat{\mathcal{U}}_{10} &= -i g e^{+\frac{i}{2}\omega_{\text{cav}} \tau} e^{-i\omega_{\text{cav}} \hat{a}^\dagger\hat{a}\tau}\mathcal{S}(\tau, \hat{\varphi}) \hat{a}^\dagger \\
        \hat{\mathcal{U}}_{11} & = e^{+\frac{i}{2}\omega_{\text{cav}} \tau} e^{-i\omega_{\text{cav}} \hat{a}^\dagger\hat{a}\tau}\biggl[ \mathcal{C}(\tau, \hat{\varphi}) +  i\frac{\Delta\omega}{2}\mathcal{S}(\tau, \hat{\varphi})\biggr]
    \end{split}
\end{equation}
in which
\begin{equation}
    \label{eq: S and C OPERATOR functions:}
\begin{split}
    \mathcal{S}(\tau, \hat{\varphi}) &= \frac{\sin \bigl( \tau\sqrt{\hat{\varphi}}\bigr) }{\sqrt{\hat{\varphi}}}  \\ 
    \mathcal{C}(\tau, \hat{\varphi}) &= \cos \bigl( \tau\sqrt{\hat{\varphi}} \bigr)
\end{split}
\end{equation}
and $\hat{\varphi}$ is associated with the number operator:
\begin{equation}
    \label{eq: varphi definition}
    \hat{\varphi} \equiv g^2 \hat{a}^\dagger \hat{a} + \bigl(\frac{\Delta\omega}{2}\bigr)^2\hat{\mathds{I}}
\end{equation}
being $\hat{\mathds{I}}$ the identity operator.

\section{The generating function of the statistics}
\label{app: the generating function}
As it was shown in Ref.~\cite{esposito2009nonequilibrium}, for a closed quantum system evolving under the unitary operator $\hat{\mathcal{U}}(\tau,0)$, it is straightforward to derive the tilted evolution operator $\hat{\mathcal{U}}_{\frac{\boldsymbol{\chi}}{2}}(\tau,0)$ via Eq.~(\ref{eq: Tilted evolution operator}). Indeed, it is sufficient to plug Eq.~(\ref{eq: Evolution operator of JC, analytical}) into it to find:
\begin{equation}
    \label{eq: Tilted U(Chi/2), explicit (appendix)}
    \hat{\mathcal{U}}_{\frac{\boldsymbol{\chi}}{2}} =  \begin{pmatrix}
        \hat{\mathcal{U}}_{00}\text{\quad\quad\quad\quad\quad\quad\quad\quad} & \hat{\mathcal{U}}_{01}e^{+i\frac{\chi_1}{2}\hbar\omega_{\text{qub}} +i\frac{\chi_2}{2}\hbar\omega_{\text{cav}} } \\
        \hat{\mathcal{U}}_{10}e^{-i\frac{\chi}{2}\hbar\omega_{\text{qub}} - i\frac{\chi_2}{2}\hbar\omega_{\text{cav}}} &  \hat{\mathcal{U}}_{11}\text{\quad\quad\quad\quad\quad\quad\quad\quad}
    \end{pmatrix}
\end{equation}
where each component $\hat{\mathcal{U}}_{ij} = \hat{\mathcal{U}}_{ij}(\tau)$ is time-dependent, as in Eq.~(\ref{eq: Evolution operator of JC, analytical}), as well as $\hat{\mathcal{U}}_{\frac{\boldsymbol{\chi}}{2}} = \hat{\mathcal{U}}_{\frac{\boldsymbol{\chi}}{2}}(\tau,0)$ (for which we dropped the time-dependence, in order to enlighten the notation). We notice that the tilted evolution operator in Eq.~(\ref{eq: Tilted U(Chi/2), explicit (appendix)}) is not so different from the original one: with respect to it, indeed, the tilted operator exhibits two additional phases in the off-diagonal elements, each one depending on the characteristic frequencies $\omega_{\text{qub}}$ and $\omega_{\text{cav}}$ only.

Let us define the density matrix of the initial joint system qubit + cavity as follows:
\begin{equation}
    \label{eq: Initial density matrix JC}
    \begin{split}
     \hat{\rho}_{\text{JC}}(0) &=  \hat{\rho}_{\text{qub}}(0)  \otimes \hat{\rho}_{\text{cav}}(0)  \\
     &=  \begin{pmatrix}
      \rho_{00}(0)   & \rho_{01}(0)  \\
       \rho_{10}(0)  & \rho_{11}(0) 
    \end{pmatrix} \otimes \hat{\rho}_{\text{cav}}(0)    
    \end{split}
\end{equation}
where $\rho_{ij}(0)$ are the $ij$-components of the qubit's initial density matrix, and $\hat{\rho}_{\text{cav}}(0)$ is the initial density matrix of the cavity. 
Since now we know the exact form (\ref{eq: Tilted U(Chi/2), explicit (appendix)}) of the tilted operator $\hat{\mathcal{U}}_{\frac{\boldsymbol{\chi}}{2}}(\tau,0)$, we can employ Eq.~(\ref{eq: Generating function}) to find the explicit expression for the generating function in the most general case:

\begin{equation}
    \label{eq: Generating function GENERAL}
    \begin{split}
        &\mathcal{G}_\tau(\boldsymbol{\chi}) =  \rho_{00}(0) \langle \zeta(\hat{a}^\dagger\hat{a}+\hat{\mathds{I}})\rangle_{\text{cav}} \\
    &+ig \rho_{01}(0) e^{+i\frac{\chi_1}{2}\hbar\omega_{\text{qub}} + i\frac{\chi_2}{2}\hbar\omega_{\text{cav}} } \langle  \hat{a}^\dagger \eta(\hat{a}^\dagger\hat{a} + \hat{\mathds{I}})\rangle_{\text{cav}} \\
    &-ig \rho_{10}(0) e^{+i\frac{\chi_1}{2}\hbar\omega_{\text{qub}} + i\frac{\chi_2}{2}\hbar\omega_{\text{cav}}} \langle  \eta^\ast(\hat{a}^\dagger\hat{a} + \hat{\mathds{I}}) \hat{a} \rangle_{\text{cav}} \\
    &+g^2 \rho_{11}(0) e^{+i\chi_1\hbar\omega_{\text{qub}} + i\chi_2\hbar\omega_{\text{cav}} } \langle \hat{a}^\dagger \mathscr{S}^2(\tau,\hat{a}^\dagger\hat{a} + \hat{\mathds{I}}) \hat{a}\rangle_{\text{cav}} \\
    &+g^2 \rho_{00}(0) e^{-i\chi_1\hbar\omega_{\text{qub}} - i\chi_2\hbar\omega_{\text{cav}} } \langle \hat{a}\mathscr{S}^2(\tau,\hat{a}^\dagger\hat{a}) \hat{a}^\dagger\rangle_{\text{cav}} \\
    &-ig \rho_{01}(0) e^{-i\frac{\chi_1}{2}\hbar\omega_{\text{qub}} - i\frac{\chi_2}{2}\hbar\omega_{\text{cav}}} \langle \eta(\hat{a}^\dagger\hat{a}) \hat{a}^\dagger\rangle_{\text{cav}} \\
    &+ig \rho_{10}(0) e^{-i\frac{\chi_1}{2}\hbar\omega_{\text{qub}} - i\frac{\chi_2}{2}\hbar\omega_{\text{cav}}}  \langle \hat{a}\eta^\ast(\hat{a}^\dagger\hat{a}) \rangle_{\text{cav}} \\
    &+\rho_{11}(0) \langle\zeta(\hat{a}^\dagger\hat{a}) \rangle_{\text{cav}}
    \end{split}
\end{equation}
where we defined: 
\begin{equation}
\label{eq: Definition of zeta, eta and eta star}
\begin{split}
    \zeta(\hat{a}^\dagger\hat{a}) &\equiv  \mathcal{C}^2(\tau, \hat{\varphi}) + (\frac{\Delta\omega}{2})^2\mathcal{S}^2(\tau, \hat{\varphi})  \\
      \eta(\hat{a}^\dagger\hat{a})  &\equiv \mathcal{S}(\tau, \hat{\varphi})\biggl(\mathcal{C}(\tau, \hat{\varphi}) -i \frac{\Delta\omega}{2}\mathcal{S}(\tau, \hat{\varphi})\biggr)  \\
     \eta^\ast(\hat{a}^\dagger\hat{a}) &\equiv  \mathcal{S}(\tau, \hat{\varphi})\biggl(\mathcal{C}(\tau, \hat{\varphi}) +i \frac{\Delta\omega}{2}\mathcal{S}(\tau, \hat{\varphi})\biggr)  
\end{split} 
\end{equation}
and $ \langle \hat{A} \rangle_{\text{cav}} \equiv \text{Tr}_{\text{cav}}\bigl\{\hat{A}(\tau)\hat{\rho}_{\text{cav}}(0)\bigr\} $. In our notation, $f(\hat{a}^\dagger\hat{a} + \hat{\mathds{I}})$ means that the substitution $\hat{\varphi}\mapsto\hat{\varphi} + g^2 \hat{\mathds{I}}$ must be performed into definitions (\ref{eq: Definition of zeta, eta and eta star}).

Expression (\ref{eq: Generating function GENERAL}) greatly simplifies if we choose, for example, $\hat{\rho}_{\text{qub}}(0) = \ket{g}\bra{g}$: by adopting the convention $\ket{e} = \footnotesize \begin{pmatrix} 1 \\ 0 \end{pmatrix}$ and $\ket{g} = \footnotesize \begin{pmatrix} 0 \\ 1 \end{pmatrix}$, indeed, $\rho_{ij}(0) = 1 \iff i=j=1$, otherwise $\rho_{ij}(0) = 0$ for $i\neq j$ or $i=j=0$. This is actually the case of the charging protocol of the JC quantum battery, where the qubit starts in the ground state. Now, let we further assume that the initial state of the cavity is a Fock state $\hat{\rho}_{\text{cav}}(0) = \ket{N}\bra{N}$ with $N$ photons, and let us set $\chi_2 = 0$ (for example, because we may be interested in computing the statistics of $\Delta U_\tau$ only). In that case, we get Eq.~(\ref{eq: Generating function, particular case}). This can be easily seen from Eq.~(\ref{eq: Generating function GENERAL}), which becomes 
\begin{equation}
    \label{eq: Generating function with a few assumptions, APPENDIX}
    \begin{split}
        \mathcal{G}_\tau(\boldsymbol{\chi}) &= g^2 e^{+i\chi_1\hbar\omega_{\text{qub}} } \langle \hat{a}^\dagger \mathscr{S}^2(\tau,\hat{a}^\dagger\hat{a}) \hat{a} \rangle_{\text{cav}} \\
        &\quad + \langle\zeta(\hat{a}^\dagger\hat{a}) \rangle_{\text{cav}}
    \end{split}
\end{equation}
because the only non-vanishing factor is $\rho_{11}(0)$. The operator-valued expression $\zeta(\hat{a}^\dagger\hat{a})$, defined in Eq.~(\ref{eq: Definition of zeta, eta and eta star}), as well as the expression $\hat{a}^\dagger \mathscr{S}^2(\tau,\hat{a}^\dagger\hat{a}) \hat{a}$, are then traced over the cavity's degrees of freedom, together with the initial state of the cavity. For what concerns $\zeta(\hat{a}^\dagger\hat{a})$, we have $\langle\zeta(\hat{a}^\dagger\hat{a}) \rangle_{\text{cav}} = \text{Tr}\{\zeta(\hat{a}^\dagger\hat{a}) \ket{N}\bra{N}\}$, that translates into $\bra{N}\zeta(\hat{a}^\dagger\hat{a})\ket{N}$. It is then straightforward to evaluate $\bra{N}\mathcal{S}^2(\tau, \hat{\varphi})\ket{N} = \mathscr{S}^2(\tau)$ and $\bra{N}\mathcal{C}^2(\tau, \hat{\varphi})\ket{N} = \mathscr{C}^2(\tau)$, where the functions $\mathscr{S}(\tau)$ and $\mathscr{C}(\tau)$ are respectively defined as:
\begin{equation}
      \label{eq: S(t) and C(t) functions, APPENDIX}
\begin{split}
    \mathscr{S}(\tau) &= \frac{\sin \biggl( \tau\sqrt{g^2 N + (\frac{\Delta\omega}{2})^2 } \biggr) }{\sqrt{g^2 N + (\frac{\Delta\omega}{2})^2 }} \\ 
    \mathscr{C}(\tau) &= \cos \biggl( \tau\sqrt{g^2 N + (\frac{\Delta\omega}{2})^2 } \biggr)
\end{split}
\end{equation}
Similarly, we can directly prove that the term $\langle \hat{a}^\dagger \mathscr{S}^2(\tau,\hat{a}^\dagger\hat{a}) \hat{a}\rangle_{\text{cav}}$ reads $N\mathscr{S}^2(\tau)$. 

Naturally, one is not obliged to make these assumptions: in principle, any initial state of the qubit and the cavity can be employed, and we can set $\chi_1 = 0$ if we look at the statistics of $Q_\tau$, or even maintain $\chi_1, \chi_2 \neq 0$ is we need the statistics of $W_\tau$. In the case of a cavity whose state is different from the Fock one, the results are complicated by the presence of multiple summations over the cavity energy levels. However, in that case, it is possible to proceed by numerical computation to obtain approximated results (e.g., the ones shown in Fig.~\ref{fig: FIG2} for thermal, coherent and squeezed coherent cavity states).

\section{Relations between the moments and the cumulants: sketch of proof}
\label{app: Sketch of proof}
Here we present the sketch of the proof for the general relations given in Eqs.~\ref{eq: SIMPLE RELATIONS}. The first step consists of computing the $n$-th moment of $\Delta U_\tau$, without assuming any initial state. From (\ref{eq: U, Q, W}), we get:
\begin{equation}
    \label{eq: n-th moment of Delta U}
    \begin{split}
    \langle\Delta U_\tau^n\rangle &= (-i)^n    \bigl[ ig \rho_{01}(0)     (+\frac{i}{2}\hbar\omega_{\text{qub}})^n  \langle  \hat{a}^\dagger\eta(\hat{a}^\dagger\hat{a} + \hat{\mathds{I}})  \rangle_{\text{cav}} \\
    &-ig \rho_{10}(0)     (+\frac{i}{2}\hbar\omega_{\text{qub}})^n \langle  \eta^\ast(\hat{a}^\dagger\hat{a}+ \hat{\mathds{I}}) \hat{a}\rangle_{\text{cav}} \\
    &+g^2 \rho_{11}(0)     (+i\hbar\omega_{\text{qub}})^n \langle \hat{a}^\dagger \mathscr{S}^2(t,\hat{a}^\dagger\hat{a} + \hat{\mathds{I}}) \hat{a} \rangle_{\text{cav}} \\
    &+g^2 \rho_{00}(0)     (-i\hbar\omega_{\text{qub}})^n  \langle \hat{a}\mathscr{S}^2(t,\hat{a}^\dagger\hat{a}) \hat{a}^\dagger\rangle_{\text{cav}} \\
    &-ig \rho_{01}(0)     (-\frac{i}{2}\hbar\omega_{\text{qub}})^n  \langle \eta(\hat{a}^\dagger\hat{a}) \hat{a}^\dagger\rangle_{\text{cav}} \\
    &+ig \rho_{10}(0)     (-\frac{i}{2}\hbar\omega_{\text{qub}})^n   \langle \hat{a}\eta^\ast(\hat{a}^\dagger\hat{a}) \rangle_{\text{cav}} \bigr] \\
    &\equiv \bigl(\hbar \omega_{\text{qub}}\bigr)^n F(\tau,n)
    \end{split}
\end{equation}
in which the function $F(\tau,n)$ summarizes the complicated expression containing the elements of the initial qubit density matrix, the coefficients, and the traces over the cavity operators. Analogously, we can compute the $n$-th moment of $Q_\tau$ via Eq.~(\ref{eq: U, Q, W}):
\begin{equation}
    \label{eq: n-th moment of Q}
    \langle Q_\tau^n\rangle \equiv \bigl(\hbar \omega_{\text{cav}}\bigr)^n F(\tau,n)
\end{equation}
which is evidently linked to Eq.~(\ref{eq: n-th moment of Delta U}) by a factor $(\omega_{\text{cav}} / \omega_{\text{qub}})^n$. This proves the first relation. 
\\
The second one comes straightforwardly from the binomial formula:
\begin{equation}
\label{eq: Binomial formula}
    (A + B)^n = \sum_{k=0}^{n}\begin{pmatrix}
        n \\
        k
    \end{pmatrix} A^{n-k} B^k
\end{equation}
where we set $A = Q_\tau$ and $B = -\langle Q_\tau\rangle$. 
\\
Then, in order to prove the third relation we just notice that:
\begin{equation}
    \label{eq: expectation of QU}
    \langle \Delta U_\tau^\alpha Q_\tau^{\beta}  \rangle = (-i)^{\alpha+\beta} \frac{\partial^{\alpha+\beta}\mathscr{G}_\tau(\chi_1, \chi_2)}{\partial\chi_1^\alpha \partial\chi_2^\beta}\biggl\rvert_{\chi_1 = 0 = \chi_2}
\end{equation}
Together with this, we employ the first relation, $\langle Q_\tau^n\rangle \equiv \bigl(\omega_{\text{cav}} / \omega_{\text{qub}}\bigr)^n \langle \Delta U_\tau^n\rangle$, as well as the formula (\ref{eq: Binomial formula}). 
\\
Finally, for what concerns the fourth formula, since we defined $W_\tau = \Delta U_\tau - Q_\tau$, it is necessary to use the binomial formula to compute $W_\tau^n = (\Delta U_\tau -Q_\tau)^n$ a first time; here, the substitution consists of setting $A = \Delta U_\tau$ and $B = -Q_\tau$. Then the binomial formula must be applied one more time, to find $\langle (W_\tau - \langle W_\tau \rangle)^n \rangle$. In that case, we simply substitute $A = W_\tau$ and $B = -\langle W_\tau \rangle$, and finally take advantage of Eqs.~(\ref{eq: n-th moment of Q}) and (\ref{eq: expectation of QU}).

\end{document}